\documentclass[12pt,preprint]{aastex62}
\usepackage{graphicx}
\usepackage{amsmath}
\usepackage{amssymb}
\footnotesize
\newdimen\digitwidth 
\setbox0=\hbox{\rm0}
\digitwidth=\wd0
\catcode`!=\active
\def!{\kern\digitwidth}
\normalsize

\def\lapp{\ifmmode\stackrel{<}{_{\sim}}\else$\stackrel{<}{_{\sim}}$\fi}

\begin{document}
\title{Polarimetry of the eclipsing pulsar PSR~J1748$-$2446A}

\correspondingauthor{X. P. You}
\email{yxp0910@swu.edu.cn}

\author{X. P. You}
\affiliation{School of Physical Science and Technology, Southwest University, Chongqing 400715, China}
\author{R. N. Manchester}
\affiliation{CSIRO Astronomy and Space Science, PO~Box~76, Epping, NSW~1710, Australia}
\author{W. A. Coles}
\affiliation{Electrical and Computer Engineering, University of California at San Diego, La Jolla, California 92093, USA}
\author{G. B. Hobbs}
\affiliation{CSIRO Astronomy and Space Science, PO~Box~76, Epping, NSW~1710, Australia}
\author{R. Shannon}
\affiliation{Centre for Astrophysics and Supercomputing, Swinburne University of Technology, Hawthorn, VIC 3122, Australia}

\begin{abstract}
Observations with the Parkes radio telescope of the eclipsing
millisecond binary pulsar PSR~J1748$-$2446A, which is in the globular
cluster Terzan 5, are presented. These include the first observations
of this pulsar in the 3\,GHz frequency band, along with simultaneous
observations in the 700\,MHz band and new observations around
1400\,MHz.  We show that the pulsar signal is not eclipsed in the
3\,GHz band and observe the known eclipses in the lower frequency
bands.  We find that the observed pulse signal becomes depolarized
during particular orbital phases and postulate that this
depolarization occurs because of rotation-measure fluctuations
resulting from turbulence in the stellar wind responsible for the
eclipses.
\end{abstract}


\section{Introduction}
The pulsar PSR~J1748$-$2446A (PSR~B1744$-$24A) was discovered by
\citet{lmd+90} using the 64-m diameter Parkes radio telescope. This
was the second known binary pulsar system for which the pulsed signal
is eclipsed during part of the orbit and it is the prototype for the
class of eclipsing binary pulsars now known as ``redbacks''
\citep{rob13}.  The pulsar has a pulse period of 11.56~ms and is
located within the globular cluster Terzan 5 (hence also known as
Ter\,5A) and is in a low-eccentricity orbit with an orbital period of
only 109 minutes.  The companion mass is estimated to be at least
0.089$\rm M_\odot$. The projected semi-major axis of this system (0.12
light seconds) implies that the separation between the pulsar and its
white dwarf companion is around 0.85$\rm R_\odot$.

The duration of the eclipse is variable in time and also changes as a
function of the observing frequency.  In the 20\,cm (1.4\,GHz)
observing band the eclipse duration is more than one-third of the
orbital period \citep{lmd+90}. The expected eclipse duration due to a
Roche lobe of estimated radius 0.15$\rm R_\odot$ \citep{nt92} is only
six minutes, much less than the typical eclipse duration.  Therefore,
the eclipse must be caused by an ablated wind from the
companion. \citet{nttf90} observed the pulsar at frequencies between
0.8 and 1.66\,GHz and found that the eclipse duration decreases with
observing frequency $\nu$ with a dependence of $\nu^{-0.63\pm0.18}$.

Understanding the eclipse mechanism can provide the opportunity to
probe the mechanism of the pulsar radiation and also the interaction
between the pulsar wind and the companion. There have been many
attempts to explain the eclipse mechanism, for example, refractive
effects \citep{pebk88}, pulse smearing \citep{rt91b}, cyclotron and
synchrotron absorption \citep{eic91,tbep94}, induced scattering
\citep{lm95} and free-free absorption \citep{wc88,rst89,rst91}.
However, the published mechanisms based on refraction cannot explain
the frequency dependence of the eclipse duration. Pulse smearing
models requires a longer dispersion time delay near the eclipse than
is observed \citep{nttf90}.  Cyclotron and synchrotron absorption
models require very high magnetic fields and pulsar winds that are
much hotter than is usually assumed \citep{tbep94}.  Free-free
absorption can explain the dependence between the eclipse duration and
the observing frequencies and currently seems consistent with
observations, although it does require a relatively low wind
temperature \citep{fg92}.

In this paper, we present multi-frequency observations of
PSR~J1748$-$2446A made with the Parkes 64-m radio telescope.  Our
observations and analysis methods are described in
Section~\ref{sec:data}. In Section~\ref{sec:results} we investigate
how the pulsar's polarization properties and its rotation measure (RM)
vary as function of orbital phase and we discuss the implications of
this work in Section~\ref{sec:depoln}. We summarize our conclusions in
Section~\ref{sec:concl}.

After completing most of this work, we became aware that Anna
  Bilous in her PhD thesis \citep{bil12} had also observed the
  orbital-phase dependence of polarization for PSR~J1748$-$2446A in
  bands centered at 820~MHz and 1500~MHz using the Green Bank
  Telescope. Her results show a similar radio-frequency and
  orbital-phase dependence for the linear polarisation to those
  presented here. She also attributed the linear depolarization to RM
  fluctuations in the circumstellar plasma.

\section{Observation and data analysis}\label{sec:data}
Our observations were made between November 2014 and January 2015,
using the Parkes 64-m diameter radio telescope. The data were taken in
three frequency bands which we label 40~cm, 20~cm and 10~cm, centered
at 728\,MHz, 1369\,MHz and 3100\,MHz, respectively.  In the 20\,cm
band the central beam of the multibeam receiver was used.
Observations in the 10\,cm and 40\,cm bands were simultaneous and used
the dual-band 10\,cm/50\,cm receiver.  We recorded a bandwidth of
64\,MHz at 40\,cm, 256\,MHz at 20\,cm and 1024\,MHz at 10\,cm, divided
into 128, 1024 and 1024 frequency channels for the three bands
respectively. The signal processing systems used were the Parkes
digital filterband system, PDFB4, for the 20\,cm and 10\,cm
observations, and a coherent dedispersion system, CASPSR \citep{vb11},
for the 40\,cm observations.  All of the data were recorded in full
polarization mode with 1-min subintegrations. Before each observations
of the pulsar, a 2-min pulsed calibration signal was recorded. We used
measurements of Hydra A to calibrate the flux density scale for our
pulsar observations.

All the observations were initially processed and calibrated using
{\sc psrchive} \citep{hvm04}. In brief, we removed
5\% of the band-edges along with narrow band interference.  We then
calibrated our observations using the polarisation and flux density
calibrators.  The 20\,cm observations are affected by cross-coupling
within the feed. We used the \textsc{pcm} method to account for this
cross-coupling \citep{van04c}.

We aim to investigate the properties of the pulsar as function of
orbital phase.  We therefore obtained relatively high signal-to-noise
ratio (S/N) polarization profiles at a range of orbital phases by
forming profiles for every 3 minutes of observing time. To form the
mean polarization profiles, the Faraday rotation across the observing
band must be taken into account.  Since the rotation measure (RM)
could vary as a function of orbital phase, we measured the RM for each
3-minute profile using the \textsc{psrchive} routine
\textsc{rmfit}. We initially searched over a large range in RM (from
$-$1000 to 1000\,rad\,m$^{-2}$) and determined the RM which maximized
the linearly polarized intensity.  To improve on this estimate the
total observing band was divided into two parts and the RM value was
used to sum the polarization profile in each half.  The final RM value
was then obtained by calculating the weighted mean position angle
difference between the two bands and, if necessary, iterating.

\section{Results}\label{sec:results}

\subsection{Variation of flux density with orbital phase}
\begin{figure*}
 \includegraphics[width=13cm,angle=-90]{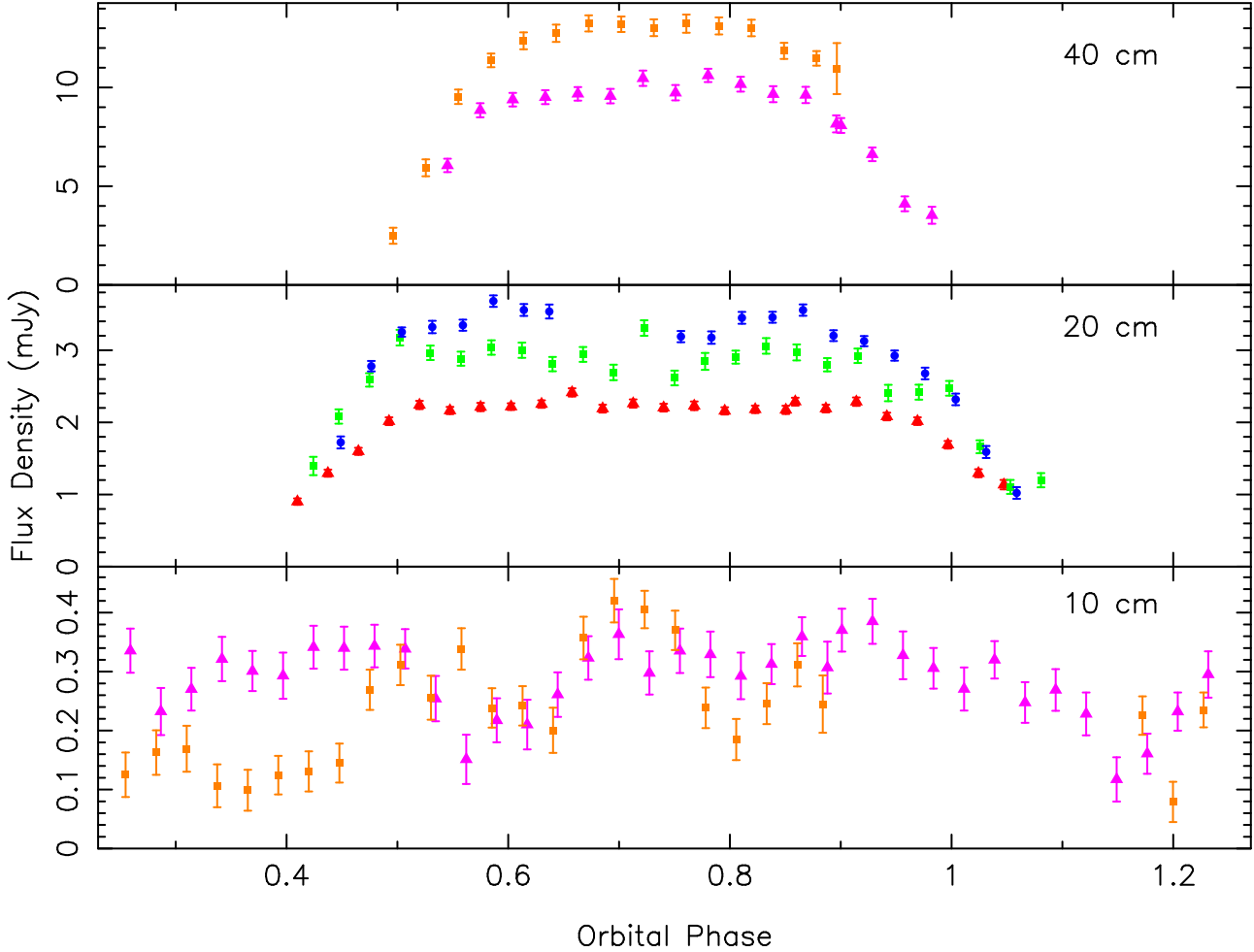}
 \caption{Mean pulse flux density as a function of orbital phase. The
   upper, middle and lower panels show the results for observations at
   40\,cm, 20\,cm, and 10\,cm respectively, with different colors
   representing different observation dates.  Observations were made
   simultaneously in the 10\,cm and 40\,cm bands and so we use the
   same color for these two bands. Observation dates (MJD) were:
   red: 56973.27, green: 57003.92, blue: 57007.21, purple: 57032.12, orange: 57041.14,
   Note that for two
   observations there are some missing points corresponding to gaps in
   the observations (blue: phases 0.63 -
   0.75, orange: phases 0.9 - 1.17).}
  \label{fg:flux}
\end{figure*}

The mean pulsed flux density ($S$) as function of orbital phase for
the three observing bands is presented in Figure~\ref{fg:flux} with
the independent observations shown in different colors.  It is clear
that the duration of the eclipse decreases with increasing observation
frequency as already found by \citet{lmd+90} and \citet{nttf90}.  We
find no evidence of any eclipse in the 10\,cm band (bottom panel of
Figure~\ref{fg:flux}) although the observed flux density is quite
variable at all orbital phases.

Although \citet{lmd+90} found the eclipse durations to be variable,
the eclipses we observed in a given band were similar in
length. Figure~\ref{fg:flux} shows that the rate of egress from the
eclipse (around orbital phase 0.5) was significantly quicker than the
rate of ingress (around orbital phase 1.0) for both 20\,cm and
40\,cm. We quantified this by approximating the engress and ingress as
linear changes and measuring the phase interval between zero and
maximum flux density (or vice versa). For the 20\,cm band, the rate of
egress is variable, with phase intervals of about 0.10 for two days
(blue and green in Figure~\ref{fg:flux}) and 0.20 for the other day
(red). The rate of ingress was slower, but more stable with the
interval for all days being close to 0.21. For 40\,cm, only one egress
and one ingress is measurable, with intervals of about 0.09 and 0.16,
respectively. A similar pattern of slow ingress and more rapid egress
is seen in other eclipsing systems, e.g., PSR B1975+20 \citep{rt91b}.

\subsection{Polarization profiles}
\begin{figure*}
  \includegraphics[width=4.4cm,angle=-90]{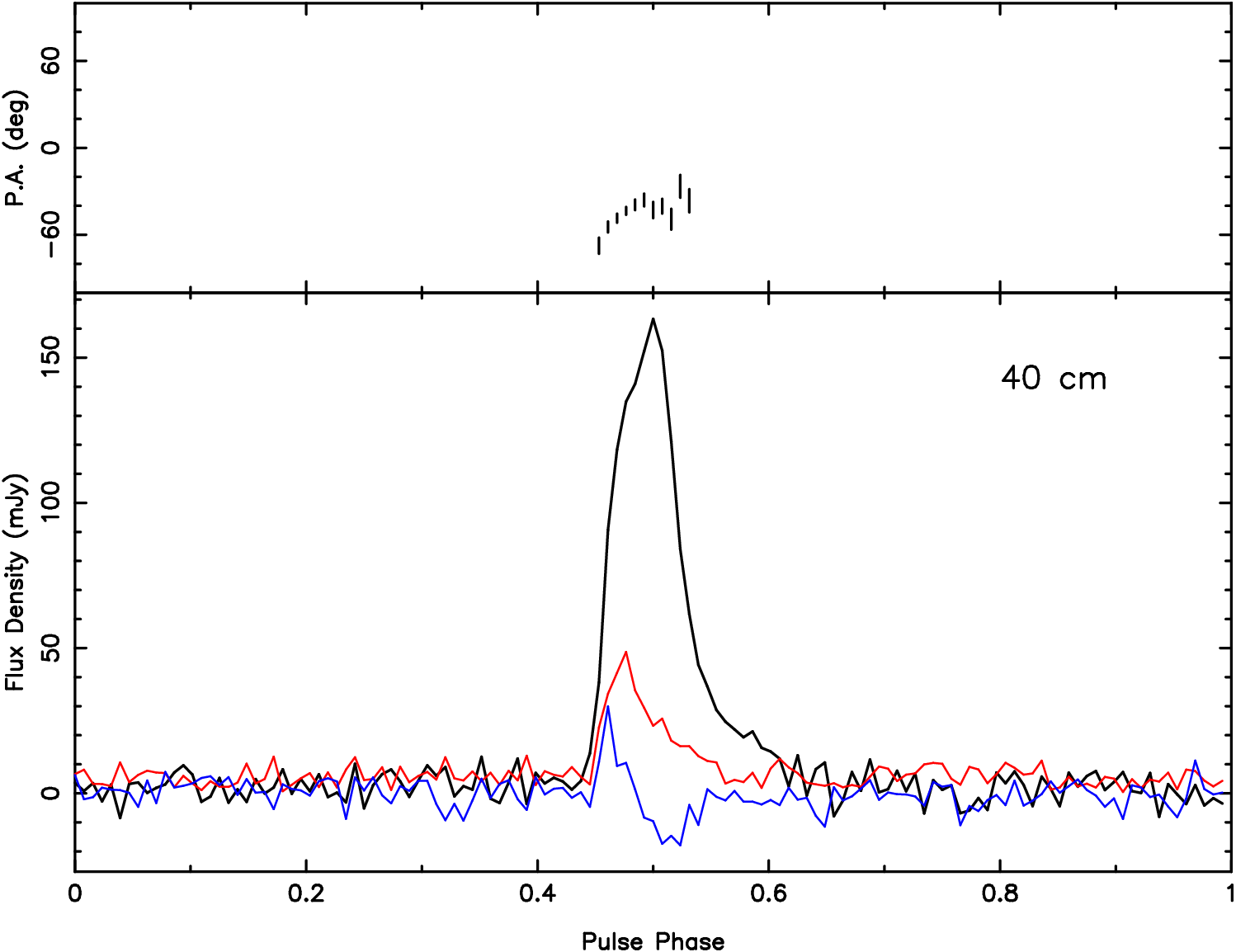}
  \includegraphics[width=4.4cm,angle=-90]{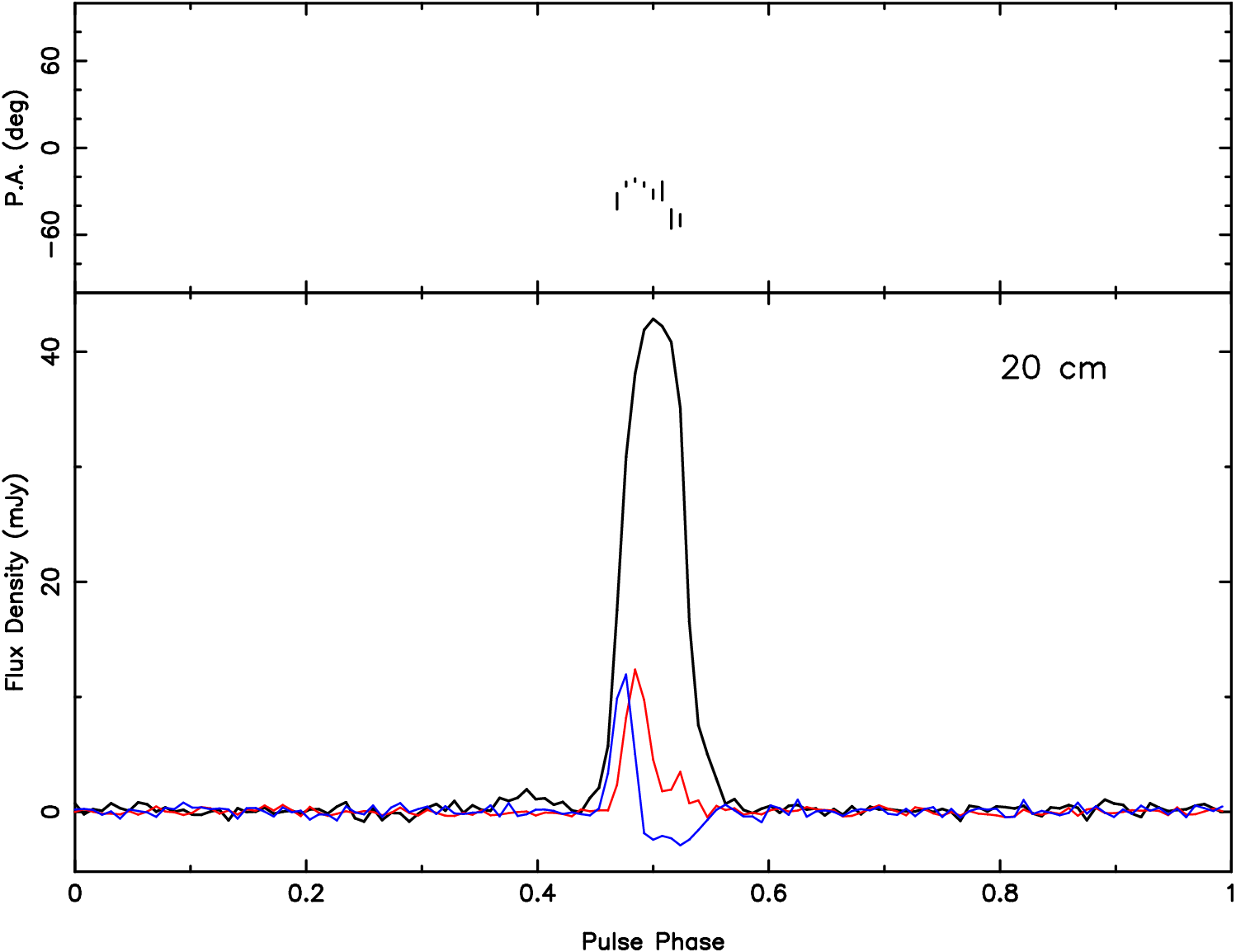}
  \includegraphics[width=4.4cm,angle=-90]{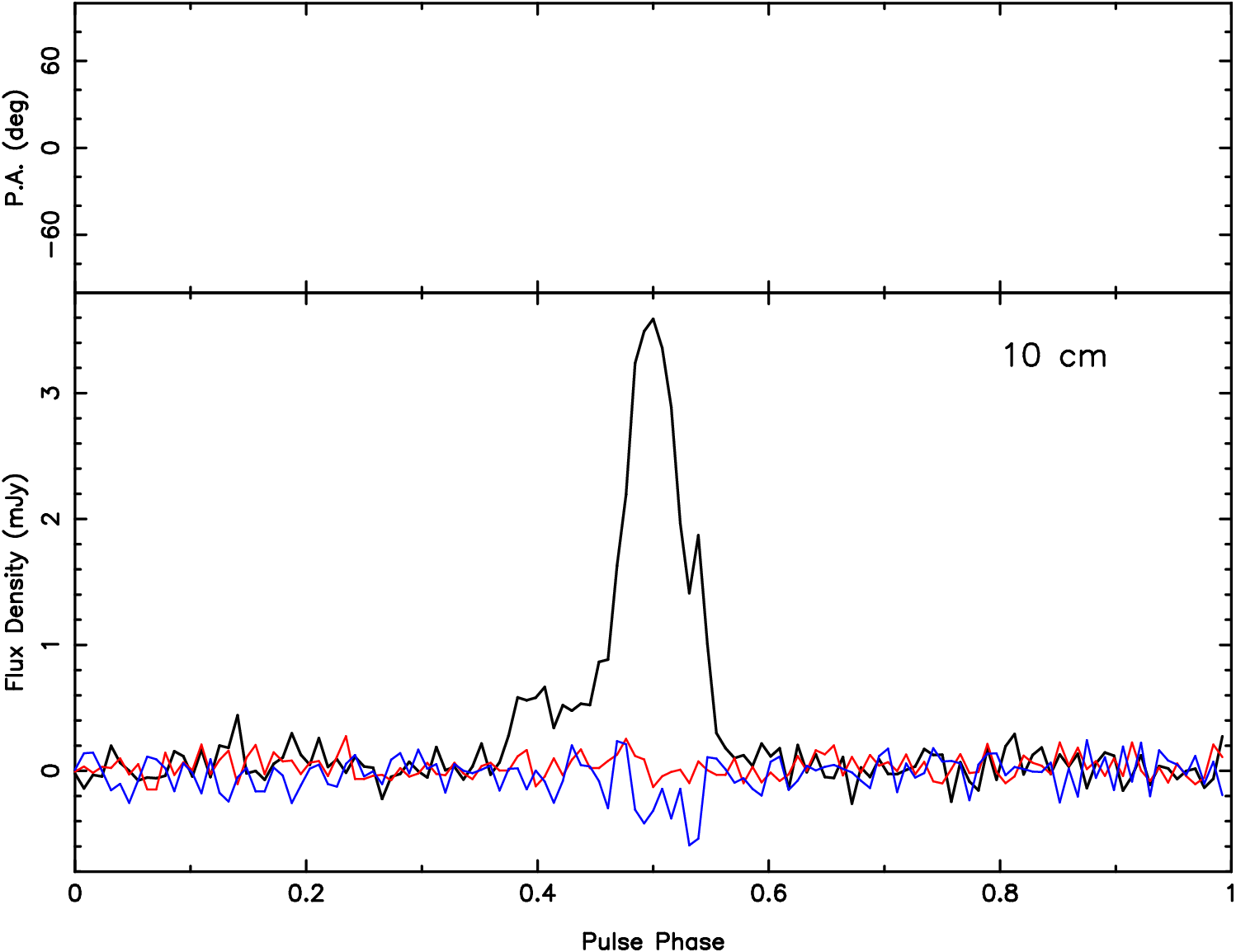}
  \caption{Mean pulse polarization profiles of PSR J1748$-$2446A in
    the 40\,cm, 20\,cm and 10\,cm observing bands. The lower panels
    give the total intensity (Stokes $I$, black line), linearly polarized
    intensity ($L = (Q^2 + U^2)^{1/2}$, where $Q$ and $U$ are the
    linear Stokes parameters, red line) and circularly polarized
    intensity (Stokes $V = I_{\rm LH} - I_{\rm RH}$, blue line)
    \citep[cf.][]{vmjr10}. The upper panels give the position angle of the
      linear polarization. Profile alignment is not absolute; the
      profile peaks are simply placed at pulse phase 0.5. }\label{fg:poln}
\end{figure*}

To measure the mean profile polarization in each band, we used the
following procedure. We first measured the RM for every 3-minute
profile.  As will be discussed further in Section~\ref{sec:depoln}
below, we found that the RM was measurable only for profiles in the
central part of the non-eclipsed phase at 20\,cm and 40\,cm. The RM
was found to be stable throughout these central phase ranges and so we
were able to sum those 3-minute profiles to form a high S/N profile
for each band as shown in Figure~\ref{fg:poln}.  We then redetermined
the RM using these summed profiles, obtaining values of 186.4$\pm$6.2
rad m$^{-2}$ at 40\,cm and 174.9$\pm$4.2 rad m$^{-2}$ at 20\,cm. These
two values are consistent within their uncertainties. The weighted
mean RM, 178.5$\pm$3.5 rad m$^{-2}$, was used to make the plots shown in
Figure~\ref{fg:poln}.

It is evident that the pulse total intensity profiles evolve strongly
across the three bands with a prominent leading
component seen at 10\,cm. The relative alignment of the
profiles in pulse phase is somewhat uncertain, but the circular
polarization profiles suggest that the main component at 10\,cm aligns
with the main component at the two lower bands. It is likely that some of
the signal on the trailing edge of the 40\,cm profile results from
interstellar scattering \citep[cf.,][]{nt92}. Never-the-less, it
appears that the leading part of the profile has a flatter spectrum
than the trailing part.

We did not detect any linear polarization at any orbital phase for the
10\,cm band, either using the 3-minute profiles or using the summation
of all the observations.  Since we see no evidence for an eclipse in
this band, the low level of linear polarization at 10\,cm or 3\,GHz is
likely to be intrinsic to the emission mechanism. For all three
observing bands we measure significant circular polarization (shown
with a blue line in the Figure) with a similar variation across the
profile in all three bands, although at 10\,cm the leading
  positive peak is barely significant.

Since the position angles in Figure~\ref{fg:poln} are calibrated
according to the IAU convention \citep{vmjr10} and the pulse profiles
and PA variations across the 40\,cm and 20\,cm profiles are similar
and relatively flat, we can use the weighted mean position angle
difference between these bands to obtain an improved estimate of the
RM. The derived value is 181.2$\pm$0.5 rad m$^{-2}$. The small
differences in the form of PA variations across the 40\,cm and 20\,cm
profiles almost certainly are intrinsic to the pulse emission, but
never-the-less result in a small variation in apparent RM across the
profile. The quoted uncertainty on the mean RM includes the effect of
this variation.

\subsection{Depolarization}\label{sec:depoln}
\begin{figure*}
  \includegraphics[width=8.0cm]{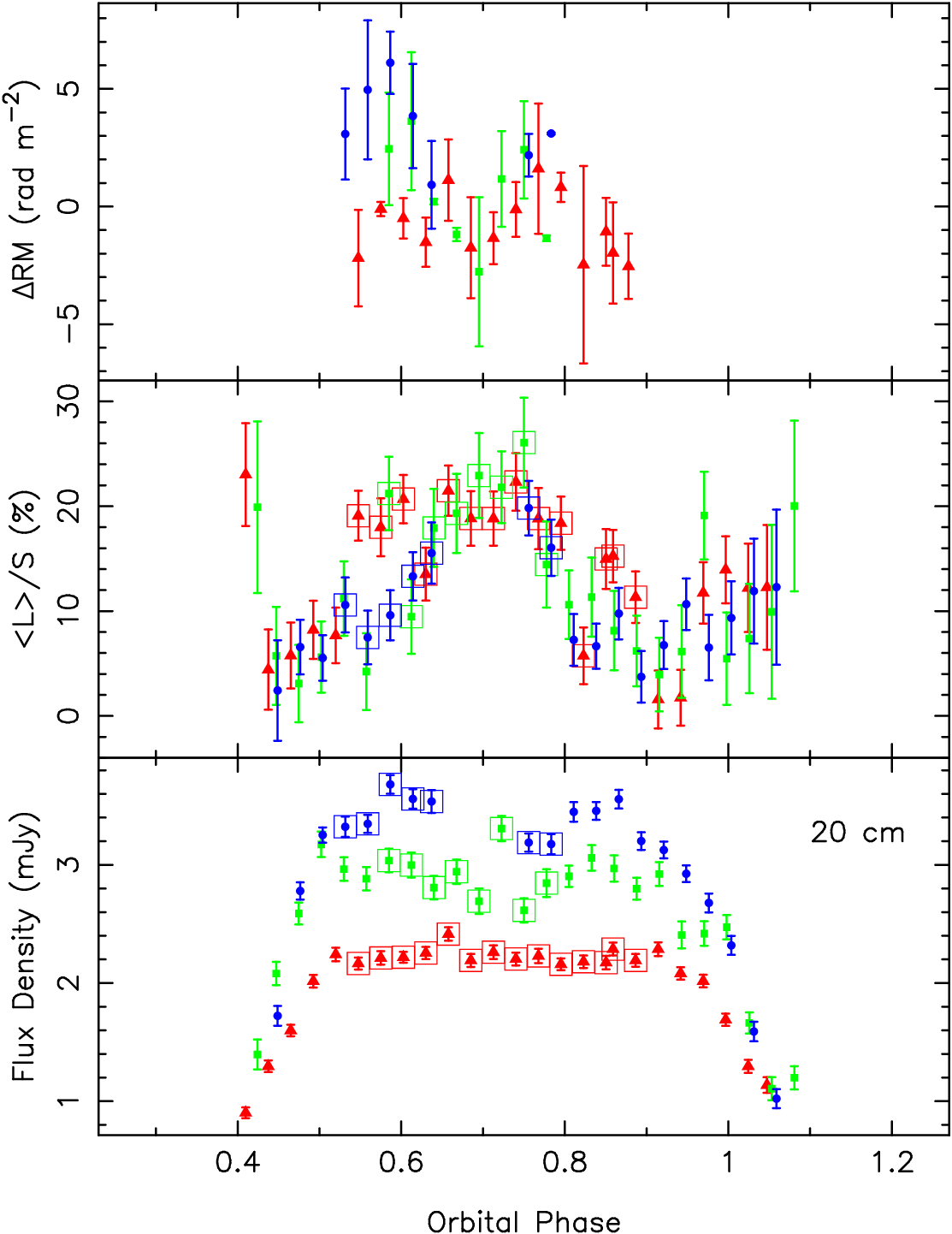}
  \includegraphics[width=8.0cm]{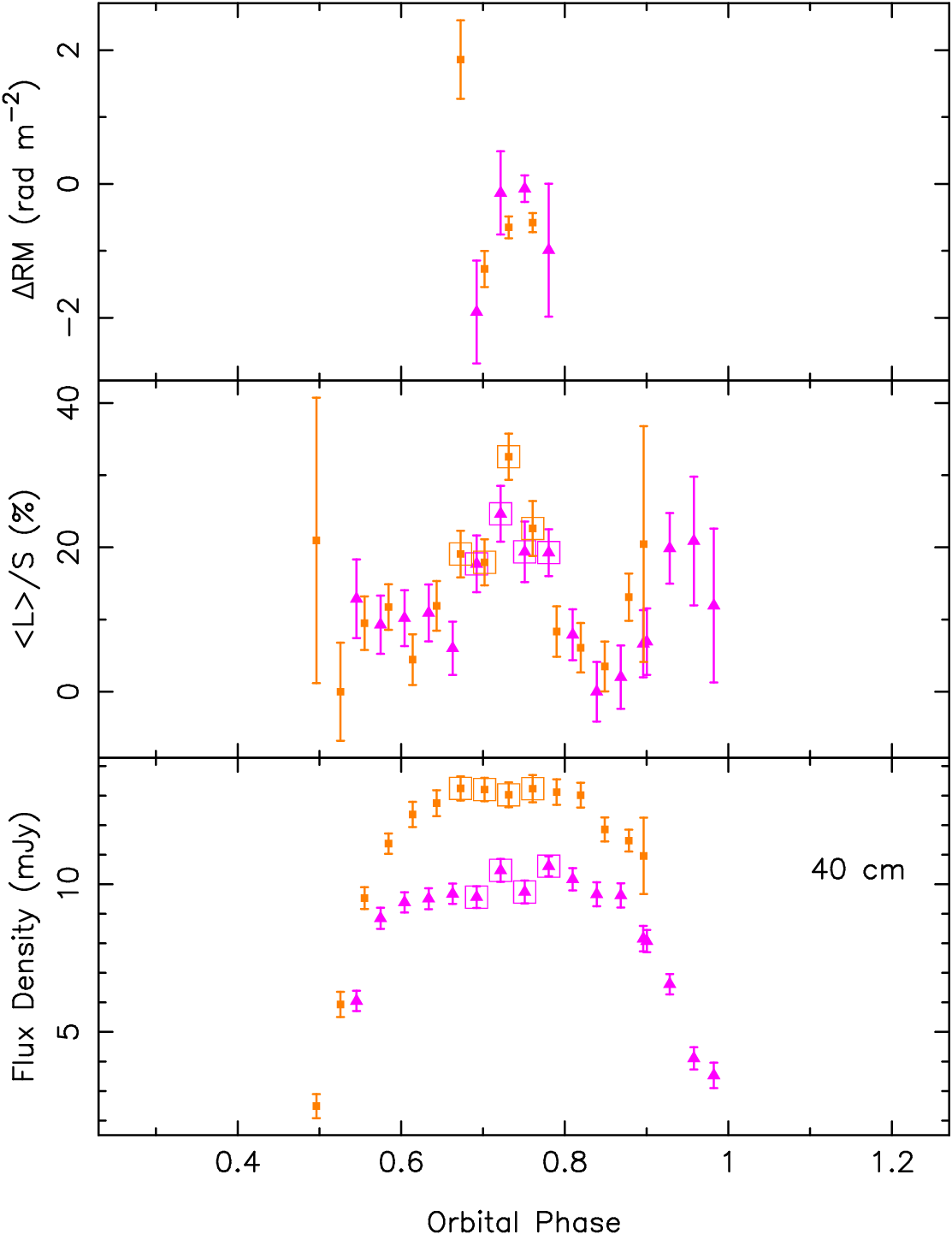}
  \caption{Polarization properties of PSR~J1748$-$2446A at 20\,cm and
    40\,cm as a function of orbital phase. The bottom panels for each
    band show the pulse mean flux density for each observing frequency
    as a function of orbital phase. As in Figure~\ref{fg:flux}, the
    different colours represent different observation dates. The
    middle panels show the mean fractional linear polarization
    $\langle L\rangle / S$, where $L=\sqrt{Q^2 + U^2}$ and $Q$ and $U$
    are the linear Stokes parameters. Since $L$ is a positive-definite
    quantity, it has a non-zero mean from system noise even when the
    pulse is effectively unpolarized. Points marked with squares
    indicate that the pulse is significantly linearly polarized at
    that orbital phase. The upper panels show the RM variations based
    on position angle differences relative to the mean profiles shown
    in Figure~\ref{fg:poln}.}
  \label{fg:depoln}
\end{figure*}

Figure~\ref{fg:depoln} shows the variation of pulsed flux density,
mean fractional linear polarization and RM as a function of orbital
phase for the 40\,cm and 20\,cm bands. We can divide the orbit into
three zones: a) when the pulse is fully eclipsed , b) when the pulse is
detectable but unpolarized and c) when the pulse is detectable and
significantly linearly polarized. As discussed above, we see no
eclipse nor any significant linear polarization for the 10\,cm-band
data.

The well-known frequency dependence of the eclipse duration is evident
from the bottom panels. The middle panels show that phase c), when
significant linear polarization is detected, is concentrated in the
middle of the un-eclipsed region and is of significantly shorter duration
than the un-eclipsed phase. Consequently, there are zones of type b),
that is, detectable pulses with no significant linear polarization, on
either side of the type c) zone. The duration of zone
c) is about 12 minutes (0.11 in phase) at 40\,cm and 33 minutes (0.30
in phase) at 20\,cm.  Zone b) is wider on ingress to the eclipse
(around phase 1.0) at both 20\,cm and 40\,cm, but has a total duration
(sum of egress and ingress zones) that is similar for the two bands, about
40 minutes or 0.37 in phase.

We determine the RM variations shown in the top panel of
Figure~\ref{fg:depoln} by comparing the average position angle
\begin{equation}\label{eq:pa}
  \psi_{\rm av} = 0.5 \tan^{-1}\frac{\langle U\rangle}{\langle Q\rangle},
\end{equation}
of each 3-minute profile having significant linear polarization with
the average position angle of the mean profile shown in
Figure~\ref{fg:poln}.  The Stokes parameters in Equation~\ref{eq:pa}
are averaged over the part of the pulse profile with significant $L$.
When the linear polarization is significant (zone c), the RM is
relatively stable.

\section{Discussion}

\subsection{Eclipse duration}
The eclipse durations at 20\,cm and 40\,cm illustrated in
Figures~\ref{fg:flux} and \ref{fg:depoln} are about 32\% and 50\% of
the orbital period, consistent with previous results
\citep{nttf90,tn91}. These observed durations give a frequency
dependence $\sim \nu^{-0.71\pm 0.15}$. This is consistent with the
$\nu^{-0.63\pm0.18}$ dependence derived by \citet{nttf90}. However, if we
extrapolate this result to the 10\,cm observing band then the eclipse
duration would be about 0.18 of the orbital period or about 20
minutes.  This would easily be detectable, but we do not see any
evidence for such an eclipse. Our observing bands do not provide
sufficient coverage to determine the frequency at which the eclipse
becomes undetectable, but it certainly occurs between 1.6~GHz
\citep{nttf90} and 2.6~GHz, the lower edge of our 10\,cm band.

Although no eclipses are seen at 10\,cm, Figure~\ref{fg:flux} shows
considerable variability across all orbital phases in the observed
flux density. The lack of dependence on orbital phase suggests that
this variability is not related to the circumstellar gas. It is more
likely to result from diffractive interstellar scintillation.

\subsection{Depolarization mechanism}
As Figure~\ref{fg:depoln} shows, the pulsar pulse is depolarized
around the boundaries of the eclipse, with the depolarization more
pronounced in the 40\,cm band compared to the 20\,cm band. The
  most likely cause of this depolarization is multi-path propagation
  in the circumstellar magnetised plasma leading to rapid time
  variations in the RM, although the traversal of the line of sight
  through the circumstellar plasma due to orbital motion will also
  contribute. Such rapid RM variations can result from fluctuations in
  either or both of the circumstellar component of DM and the
  parallel component of the  magnetic field. DM variations are observed at both
  eclipse ingress and egress, with those at egress generally more
  significant \citep{rt91b,nt92}. However, these variations are
  typically $\lapp 0.1$~cm$^{-3}$~pc which will not significantly
  affect the RM. There is weak evidence for broadening
  of the mean pulse profiles at eclipse ingress or egress relative to
  the uneclipsed profile at the 10\% level by $\lapp 0.5$~ms,
  corresponding to a $\Delta{\rm DM} \lapp
  0.2$~cm$^{-3}$~pc. Consequently DM fluctuations are unlikely to be
  responsible for the depolarization. Therefore small-scale variations
  in the line-of-sight component of the magnetic field are the most
  likely cause of the depolarization, irrespective of whether the RM
  variations result from multi-path propagation or orbital motion of
  the line of sight. Such fluctuations would be expected in a
  turbulent stellar wind. 

If the RM variations have a normal distribution with standard
deviation $\sigma_{\rm RM}$, the variations in polarization position
angle $\psi$ will also be normal with standard deviation $\sigma_\psi
= \lambda^2 \sigma_{\rm RM}$. Averaging over this variation will
reduce the linear polarization $L$ by a factor $\exp(-2\sigma_\psi^2)
= \exp(-2\lambda^4 \sigma_{\rm RM}^2)$ (see Appendix A).

As expected and as observed, the depolarization mechanism is more
effective at longer wavelengths. If we assume that the depolarization
is significant when $\sigma^2_{\rm RM}\lambda^4>1$, then
depolarization will occur if $\sigma_{\rm RM}$ is greater than 6\,rad
m$^{-2}$ at 40\,cm, 25\,rad m$^{-2}$ at 20\,cm and 100\,rad m$^{-2}$
at 10 cm respectively. Figure~\ref{fg:depoln} shows that, at orbital
phases from about 0.55 and 0.67 and from 0.78 to 0.90 (when the pulsar
-- companion line is between $20\degr$ and $63\degr$ from the line of
sight) the pulse is depolarized at 40\,cm but not at
20\,cm. Consequently at these orbital phases, $\sigma_{\rm RM}$ must
be between 6\,rad m$^{-2}$ and 25\,rad m$^{-2}$. The top panel of
Figure~\ref{fg:depoln} shows RM fluctuations of this order. For a
fluctuating electron density of order $10^6$~cm$^{-3}$ \citep{rst91}
and a path length of order 1~$R_\odot$, the implied magnetic field is
of order 10~$\mu$G which is plausible. More detailed modelling would
require three-dimensional models of both the circumstellar plasma and
the magnetic field structure.

\section{Conclusions}\label{sec:concl}
Polarization observations of the eclipsing binary pulsar
J1748$-$2446A, the first-discovered pulsar in the globular cluster
Terzan 5, have been made in three bands using the Parkes 64-m radio
telescope: 40\,cm (centered at 728~MHz), 20\,cm (1369~MHz) and, for
the first time, at 10\,cm (3100~MHz). Observations in the two
lower-frequency bands show the well-known eclipses, but no eclipses
are evident at 10\,cm. Away from the eclipses, the two lower-frequency
bands show significant linear and circular polarization, but at
10\,cm, there is no detectable linear polarization. As the pulsar
enters and leaves the eclipse, the linear polarization at 20\,cm and
40\,cm is quenched. We attribute this depolarization to
short-timescale RM fluctuations caused by time-variable scattering and
the resultant multi-path propagation in the circumstellar plasma, and
to the changing propagation path due to orbital motion. An
ultra-wide-bandwidth receiver, currently being commissioned at the
Parkes radio telescope, will have increased sensitivity and frequency
coverage compared to the current receivers. This new receiver will be
ideal for carrying out further studies of this pulsar and the orbital
dependence of its polarization.

\section*{Acknowledgments}

Our initial observations were based on a timing model for
PSR~J1748$-$2446A provided by S. M. Ransom and obtained using the
Green Bank radio telescope.  XPY is supported by the National Natural
Science Foundation of China (U1231120, 11573008) and the Fundamental Research
Funds for the Central Universities (XDJK2015B012).  GH is supported by
an Australian Research Council Future Fellowship grant. The data
presented in this paper were obtained by the Parkes radio telescope,
which is part of the Australia Telescope which is funded by the
Commonwealth of Australia for operation as a National Facility managed
by CSIRO.

\software{psrchive \citep{hvm04},
  https://sourceforge.net/projects/psrchive/​, version 2012-09+}


\begin{thebibliography}{18}
\expandafter\ifx\csname natexlab\endcsname\relax\def\natexlab#1{#1}\fi

\bibitem[{Bilous(2012)}]{bil12}
Bilous, A. 2012, PhD thesis, University of Virginia

\bibitem[{Born \& Wolf(1970)}]{bw70}
Born, M. \& Wolf, E. 1970, Principles of Optics, Cambridge Press

\bibitem[{Eichler(1991)}]{eic91}
Eichler, D. 1991, ApJ, 370, L27

\bibitem[{Fruchter \& Goss(1992)}]{fg92}
Fruchter, A.~S. \& Goss, W.~M. 1992, ApJ, 384, L47

\bibitem[{{Hotan} {et~al.}(2004){Hotan}, {van Straten}, \&
  {Manchester}}]{hvm04}
{Hotan}, A.~W., {van Straten}, W., \& {Manchester}, R.~N. 2004, PASA, 21, 302

\bibitem[{Luo \& Melrose(1995)}]{lm95}
Luo, Q. \& Melrose, D.~M. 1995, ApJ, 452, 346

\bibitem[{Lyne {et~al.}(1990)Lyne, Manchester, D'Amico, Staveley-Smith,
  Johnston, Lim, Fruchter, Goss, \& Frail}]{lmd+90}
Lyne, A.~G., Manchester, R.~N., D'Amico, N., Staveley-Smith, L., Johnston, S.,
  Lim, J., Fruchter, A.~S., Goss, W.~M., \& Frail, D. 1990, Nature, 347, 650

\bibitem[{Nice \& Thorsett(1992)}]{nt92}
Nice, D.~J. \& Thorsett, S.~E. 1992, ApJ, 397, 249

\bibitem[{Nice {et~al.}(1990)Nice, Thorsett, Taylor, \& Fruchter}]{nttf90}
Nice, D.~J., Thorsett, S.~E., Taylor, J.~H., \& Fruchter, A.~S. 1990, ApJ, 361,
  L61

\bibitem[{Papoulis, A.(1991)}]{pap91}
Papoulis, A. 1991, Probability, Random Variables, and Stochastic
Processes, McGraw Hill

\bibitem[{Phinney {et~al.}(1988)Phinney, Evans, Blandford, \&
  Kulkarni}]{pebk88}
Phinney, E.~S., Evans, C.~R., Blandford, R.~D., \& Kulkarni, S.~R. 1988,
  Nature, 333, 832

\bibitem[{Rasio {et~al.}(1989)Rasio, Shapiro, \& Teukolsky}]{rst89}
Rasio, F.~A., Shapiro, S.~L., \& Teukolsky, S.~A. 1989, ApJ, 342, 934

\bibitem[{Rasio {et~al.}(1991)Rasio, Shapiro, \& Teukolsky}]{rst91}
---. 1991, A\&A, 241, L25

\bibitem[{{Roberts}(2013)}]{rob13}
{Roberts}, M.~S.~E. 2013, in IAU Symposium, Vol. 291, Neutron Stars and
  Pulsars: Challenges and Opportunities after 80 years, ed. J.~{van Leeuwen},
  127--132

\bibitem[{{Ryba} \& {Taylor}(1991)}]{rt91b}
{Ryba}, M.~F. \& {Taylor}, J.~H. 1991, ApJ, 380, 557

\bibitem[{Thompson {et~al.}(1994)Thompson, Blandford, Evans, \&
  Phinney}]{tbep94}
Thompson, C., Blandford, R.~D., Evans, C.~R., \& Phinney, E.~S. 1994, ApJ, 422,
  304

\bibitem[{Thorsett \& Nice(1991)}]{tn91}
Thorsett, S.~E. \& Nice, D.~J. 1991, Nature, 353, 731

\bibitem[{{van Straten}(2004)}]{van04c}
{van Straten}, W. 2004, ApJS, 152, 129

\bibitem[{{van Straten} \& {Bailes}(2011)}]{vb11}
{van Straten}, W. \& {Bailes}, M. 2011, PASA, 28, 1

\bibitem[{{van Straten} {et~al.}(2010){van Straten}, {Manchester}, {Johnston},
  \& {Reynolds}}]{vmjr10}
{van Straten}, W., {Manchester}, R.~N., {Johnston}, S., \& {Reynolds}, J.~E.
  2010, PASA, 27, 104

\bibitem[{Wasserman \& Cordes(1988)}]{wc88}
Wasserman, I. \& Cordes, J.~M. 1988, ApJ, 333, L91

\end{thebibliography}

\appendix
\section{Derivation of depolarization factor}
Any partial coherence such as depolarization
or reduction in visibility of an interferometer, requires both finite
bandwidth and finite integration time \citep{bw70}. Partial
polarization can be described by coherency matrix $\bf C = E E^T$
where the 2-vector $\bf E$ components are any two orthogonal
polarizations. Here we take a linear basis.  We then consider the
effect of multi-path propagation on a unit amplitude wave polarized in
the x direction. The different paths have different Faraday rotation
angle $\psi$. Then $\bf C$ becomes
\begin{equation}
  {\bf C} =\left[ {\begin{array}{cc}
    \cos^2\psi & -\cos\psi \sin\psi \\
    -\cos\psi \sin\psi & \sin^2\psi
\end{array}}\right].
\end{equation}
We assume $\psi$ to be a zero mean gaussian random variable and
  take the expectation of $\bf C$ over paths, thus explicitly over
  $\psi$ and implicitly over time. The expectations of the off
  diagonal terms $C_{xy}$ and $C_{yx}$ are zero as $\psi$ has an even
  distribution. The diagonal terms can be written
\begin{equation}
{\begin{array}{l}
  \langle C_{xx}\rangle = \langle\cos^2\psi\rangle = 0.5 (1+\langle\cos 2\psi\rangle)\\
  \langle C_{yy}\rangle = \langle\sin^2\psi\rangle = 0.5 (1-\langle\cos 2\psi\rangle).
\end{array}}
\end{equation}
$\langle {\bf C}\rangle$  can be written as the sum of an unpolarized
component $\langle {\bf C_{up}}\rangle$ and a $100\%$ linearly polarized x-component
$\langle {\bf C_{p}}\rangle$  where
\begin{equation}
{\begin{array}{l}
  \langle {\bf C_{up}}\rangle =\left[ {\begin{array}{cc}
    \langle C_{yy}\rangle & 0 \\
    0 & \langle C_{yy}\rangle
\end{array}}\right],\;\;
  \langle {\bf C_{p}}\rangle =\left[ {\begin{array}{cc}
    \langle C_{xx}\rangle - \langle C_{yy}\rangle & 0 \\
    0 & 0
\end{array}}\right].
\end{array}}
\end{equation}
Thus the amplitude of the linear polarized component has been reduced from 1.0 to
\begin{equation}
  L = \langle C_{xx}\rangle - \langle C_{yy}\rangle = \langle\cos 2\psi\rangle = \exp(-2\sigma_\psi^2)
\end{equation}
where the latter equality is applicable for normal random
  processes \citep{pap91}.

\end{document}